\documentclass[USenglish,twocolumn]{article}

\usepackage[utf8]{inputenc}
\usepackage[big]{dgruyter}
\usepackage{graphicx}
\usepackage{pgfplots}
\pgfplotsset{compat=1.18,compat/show suggested version=false}
\usepackage{booktabs}
\usepackage{url}

\begin{document}

  \articletype{Research Article{\hfill}Open Access}

  \author[1]{Serhii Yemets}

  \author*[2]{Marek Horv\'ath}

  \affil[1]{Technical University of Ko\v{s}ice, Letn\'a 9, 042 00 Ko\v{s}ice, Slovakia, E-mail: serhii.yemets@student.tuke.sk}

  \affil[2]{Technical University of Ko\v{s}ice, Letn\'a 9, 042 00 Ko\v{s}ice, Slovakia, E-mail: marek.horvath@tuke.sk}

  \title{\huge Source Code Authorship Attribution Does Not Generalize from Competitions to Classrooms}

  \runningtitle{Attribution Does Not Generalize from Competitions to Classrooms}

  \begin{abstract}
{Source code authorship attribution aims to identify the author of a program fragment from its writing style. We fine-tune the pre-trained transformer CodeBERT on three sources of data: publicly available Google Code Jam (GCJ) submissions from an open Kaggle repository, a curated GCJ archive, and institutional coursework datasets collected at a technical university. On multi-round GCJ data, CodeBERT reaches 92.6\% Top-1 accuracy for 10 authors and retains 70.7\% Top-1 (88.2\% Top-10) for 1000 authors. On the examined coursework datasets, the same pipeline performs at or below the corresponding chance baselines: 0.2\% Top-1 on a closed-assignment dataset of 690 authors and 0.06\% Top-1 on open-ended assignments evaluated over 812 authors. A companion multi-model benchmark provides consistent cross-model evidence across additional dataset configurations, indicating that the observed performance gap is not specific to CodeBERT and persists across the evaluated model families. We analyze dataset and task properties that plausibly explain this gap and argue that GCJ-based benchmarks overestimate the practical applicability of authorship attribution in educational settings unless they are validated on the target coursework context.}
\end{abstract}

  \journalname{Open Computer Science}

  \journalyear{2026}
  \journalvolume{1}
  \journalissue{1}

\maketitle

\section{Introduction}
Programmers develop individual habits in naming, indentation, decomposition, and the way they revise a solution, and some of these habits persist across the programs they write. Source code authorship attribution is the task of recovering the author of a fragment from these stylistic traces, and it has long been applied in software forensics, malware analysis, and disputes over code ownership. In education the problem has recently gained new urgency: with the spread of large language models capable of generating working solutions to programming assignments, instructors increasingly need tools that can verify whether a submission is consistent with the personal style of the student who submitted it. Authorship attribution is therefore a plausible screening technology, but only if its evidence survives the constraints of real coursework.

Classical approaches to source code attribution rely on hand-crafted stylometric features such as identifier naming conventions, indentation, layout metrics, and \textit{n}-gram frequencies [8, 9, 10]. Deep learning promised to remove the feature engineering step: neural networks learn discriminative representations of coding style directly from raw code, and reported results are impressive. Convolutional and recurrent models identify hundreds of authors with over 90\% accuracy [3], and pre-trained transformers for code such as \textit{CodeBERT} [1] provide contextual embeddings that further simplify the pipeline.

However, nearly all of these results share a common experimental foundation: the \textit{Google Code Jam} (GCJ) corpus. GCJ is attractive because thousands of programmers solve the same tasks, so a classifier cannot exploit differences in program functionality and is forced to model style. At the same time, GCJ participants are experienced, self-selected programmers who write substantial, algorithmically rich solutions under time pressure and without common instructor-provided templates. These properties make GCJ an unusually favorable environment for stylometry, and it is an open question whether accuracy measured there says anything about performance on the data instructors actually possess: coursework submitted by first- and second-year students.

This paper addresses that question directly. We evaluate the same authorship-attribution framing on three sources of data with distinct provenance. The first is a publicly available GCJ dataset retrieved from an open Kaggle repository, from which we build the large multi-round configurations used for the scaling study. The second is a curated GCJ archive based on individual contest rounds. The third comprises institutional coursework datasets collected at a technical university; a companion benchmark [17] supplies secondary evidence across additional dataset configurations and model families. The first two sources share the same competitive nature and are analyzed jointly as the competitive programming family, while the coursework data represent the educational setting targeted by the study. Our contribution is threefold. First, we replicate and extend the strong GCJ results, scaling a fine-tuned CodeBERT classifier from 10 to 1000 authors while retaining high Top-1 and Top-10 accuracy. Second, applying the same primary CodeBERT pipeline to the examined coursework datasets, we show that performance approaches or falls below the corresponding chance baselines. Third, we use the companion benchmark to test whether the same pattern is visible beyond CodeBERT, finding consistent evidence across character-, token-, recurrent, and transformer-based model families. The following research questions guide the study:

\begin{itemize}
  \item \textbf{RQ1:} How does the accuracy of a fine-tuned code transformer scale with the number of candidate authors on competitive programming data?
  \item \textbf{RQ2:} Does accuracy obtained on competitive programming corpora transfer to institutional coursework under the same primary CodeBERT pipeline?
  \item \textbf{RQ3:} Is the observed behavior specific to one architecture, or does it persist across unrelated model families?
  \item \textbf{RQ4:} Which properties of educational datasets limit the applicability of deep learning-based authorship attribution?
\end{itemize}

\section{Related Work}
Research on source code authorship attribution spans more than three decades and has passed through several methodological eras. The earliest work framed the problem as \textit{software forensics} and relied on manually designed metrics. Krsul and Spafford proposed more than fifty measurements grouped into program layout, style, and structure metrics, and reported that twenty-nine programmers could be attributed with roughly seventy-three percent accuracy [10]. This line established the central assumption of the field, namely that a programmer leaves a consistent stylistic fingerprint, while also exposing its fragility: layout features are easily erased by code formatters, and inexperienced authors have often not yet settled into a stable style.

A second generation of methods moved from hand-picked metrics to \textit{n}-gram profiles. Frantzeskou et al.\ introduced the Source Code Author Profile (SCAP) approach, which represents each author by their most frequent byte-level \textit{n}-grams and compares an unknown sample to these profiles using a simplified profile intersection measure [8]. The method is language-independent, computationally simple, and effective even when only a few short programs per author are available. Burrows et al.\ subsequently unified and compared the competing feature-based techniques on a common footing, evaluating \textit{n}-gram and metric representations with several information-retrieval ranking functions over a corpus of student programming assignments [9]. Two of their observations directly anticipate our own findings: the problem domain of the software strongly affects attribution accuracy, and a programmer's coding style drifts over time regardless of the requirements being coded.

The current dominant benchmark was established by Caliskan-Islam et al., who introduced a Code Stylometry Feature Set combining lexical, layout, and syntactic features extracted from abstract syntax trees, produced by a fuzzy parser able to handle incomplete code, and classified with random forests [2]. Their system de-anonymized 1,600 \textit{Google Code Jam} programmers with roughly ninety-four percent accuracy and, in doing so, made GCJ the de facto standard benchmark for the task [5]. The same work reported two properties that make GCJ especially favorable for stylometry: coding style is largely preserved across competition years, and solutions to more difficult problems are easier to attribute, because skilled programmers write more distinctive code.

Deep learning subsequently removed the manual feature-engineering step. Abuhamad et al.\ proposed DL-CAIS, a language-oblivious model that feeds TF-IDF-weighted code \textit{n}-grams into a deep recurrent network and reports 92-97\% accuracy at scales ranging from about 1,600 up to 8,903 GCJ authors [3]. Alsulami et al.\ instead operated directly on program structure, traversing abstract syntax trees with LSTM and bidirectional LSTM networks and attributing seventy Python authors with close to eighty-nine percent accuracy [4]. More recently, transformer models pre-trained on large code corpora, such as \textit{CodeBERT} [1], which is built on the BERT architecture [11, 13], have allowed attribution to be cast as straightforward sequence classification through fine-tuning. Character-level convolutional networks [12] offer a lightweight, from-scratch alternative that operates below the token level and is robust to out-of-vocabulary tokens and misspellings.

Alongside these advances, a critical literature has questioned whether benchmark accuracy reflects real-world performance. Kalgutkar et al.\ survey the field and note its reliance on clean, single-author datasets, observing that competition corpora such as GCJ have been criticized for their artificial setup, and that the anonymization required to share student data typically removes comments, one of the most discriminative features [5]. Dauber et al.\ show that individual small, incomplete code fragments are substantially harder to attribute, although aggregating several fragments linked to the same account restores high accuracy [7]. Most directly, Bogomolov et al.\ demonstrate that models with high accuracy on existing datasets degrade sharply when evaluated on more realistic data, and they introduce the notion of a work context to capture the project- and team-specific factors that attribution models fail to generalize across [6]. A recent programmer-attribution survey reaches the same methodological diagnosis from a broader mapping perspective: the field is dominated by closed-world attribution, a small number of benchmarks, and comparatively little work on behavioral evidence, verification, and reproducibility [15].

Our study extends this critical line to the educational domain, whose distinctive properties include immature and rapidly evolving style, shared assignment templates, and short tasks. We directly compare a fixed CodeBERT attribution pipeline across competitive-programming and institutional coursework data under a task-holdout protocol, and use companion cross-model evidence to examine whether the pattern persists beyond one architecture.

\section{Datasets}
Our experiments draw on two contrasting families of data: competitive programming submissions, which represent the favorable conditions under which deep attribution models are often evaluated, and institutional coursework, which represents the conditions instructors face in the examined courses. This section describes both, together with the contest structure and data-quality considerations that distinguish them.
\subsection{Competitive Programming Data (GCJ)}
The competitive part of our experiments draws on archived Google Code Jam submissions (2018-2022) from two sources of distinct provenance, used for two complementary purposes. The first source is the public Kaggle dataset \textit{Google Code Jam dataset} (\texttt{jur1cek/gcj-dataset}), from which we use multi-round configurations for the scaling study [14]. Being community-maintained, it is comparatively noisy, with occasional duplicate, truncated, or mislabeled entries, but it provides many solutions per author across multiple rounds. From it we build the three scaling configurations by selecting the 10, 500, and 1000 most active authors. This selection creates a favorable high-volume setting and is treated as a limitation when interpreting transfer to coursework. The second source is a curated set of additional single-round GCJ configurations: GCJ 2022 with 10 tasks per author (636 authors) and GCJ 2018 with 10 tasks per author (1000 authors). We fixed the number of solutions per author at ten deliberately, as a compromise between training cost and accuracy. In all configurations each solution is labeled with the participant's handle, and participants across a round solve identical tasks, which prevents the classifier from using task semantics as a shortcut. The multi-round and single-round configurations differ both in provenance and in per-author code volume, so their comparison is interpreted as descriptive rather than as an isolated causal test of either factor.

\subsection{Student Coursework Datasets}
The student datasets consist of coursework submissions collected at a technical university, covering introductory C programming, Java-based object-oriented tasks, and related programming coursework. Before analysis, student identifiers and repository-level identifiers were replaced with salted hashes and metadata was normalized; identifiers occurring inside the submitted source code were not modified as part of this anonymization step. This corpus belongs to a broader line of educational source-code analysis in which static analysis, code-quality feedback, and learning analytics are used to study how novice programmers develop over repeated assignments [16]. For the primary CodeBERT experiments we use two assignment-type datasets. The first is a closed-assignment dataset (690 authors, 2{,}345 submissions across four fully specified C/C++ tasks), in which each assignment is a single task with one expected solution. The second is an open-assignment dataset (812 authors, 2{,}487 submissions across five C/C++/Java tasks), consisting of free-form projects in which students choose their own design and enjoy greater freedom of individual expression. For triangulation, we also use secondary evidence from the companion benchmark [17], which includes the closed and open datasets, competition subsets (Google Code Jam and Kick Start, including pre- and post-LLM-period splits), and further coursework partitions by author count, programming language, project size, and number of files per author. Because the datasets consist of educational materials reused across course editions, the raw submissions cannot be released publicly; the anonymized version is stored on institutional infrastructure and available to researchers on reasonable request.

The two competitive sources and the student source differ in exactly the dimensions that matter for stylometry. GCJ authors are experienced volunteers producing substantial contest solutions; students are novices producing shorter, assignment-constrained programs whose style is still forming. This contrast between competitive and educational data is the object of study rather than a nuisance variable to be removed.

\begin{table}[t]
\centering
\small
\caption{Core evaluation datasets.}
\label{tab:core}
\footnotesize
\setlength{\tabcolsep}{3pt}
\begin{tabular}{@{}p{1.55cm}rrp{2.35cm}p{1.15cm}@{}}
\toprule
Dataset & Authors & Samples & Tasks/lang. & Source \\
\midrule
GCJ multi & 10--1000 & --    & multiple rounds / mixed & Kaggle \\
GCJ 2022  & 636      & 6360  & 10 tasks / mixed & Curated \\
GCJ 2018  & 1000     & 10000 & 10 tasks / mixed & Curated \\
Closed    & 690      & 2345  & 4 tasks / C/C++ & Inst. \\
Open      & 812      & 2487  & 5 tasks / C/C++/Java & Inst. \\
\bottomrule
\end{tabular}
\end{table}

\subsection{Contest Structure and Data Quality}
Because our competitive data come from two Google programming contests, Google Code Jam and Kick Start, it is worth noting how their structure shapes the resulting code. Google Code Jam was a multi-round elimination contest in which participants solved time-boxed algorithmic problems with accepted submissions judged against hidden tests. This format means that a GCJ solution is usually written by an experienced, self-selected programmer in a contest setting, often with recurring personal routines for input-output handling, templates, macros, and decomposition.

Kick Start differs in one important respect: it is a series of independent three-hour rounds held throughout the year, open to everyone with no pre-qualification and no single overall winner. Its problem format closely mirrors Code Jam [14], and in practice the two contests are similar in the kind of algorithmic tasks they pose. Both therefore yield substantial, time-constrained, single-author solutions that are favorable for stylometric analysis.

We do not treat the two contests as evidence about programmer identity in general. They are used as favorable competitive-programming baselines whose production context differs sharply from coursework. Empirically, the observed differences across companion configurations are driven more by the amount and structure of available code than by the specific contest label (Section~5.4).

Table~\ref{tab:context} summarizes the expected direction of these differences. The table is not used as a scoring instrument; rather, it makes explicit which assumptions are being carried from one data context to another when competitive-programming results are interpreted as evidence for educational use. In stylometric terms, competitive data tend to increase stable between-author variation, while coursework data often increase within-author variation and between-author similarity at the same time.

\begin{table}[t]
\centering
\footnotesize
\setlength{\tabcolsep}{3pt}
\caption{Dataset properties relevant to authorship signal.\label{tab:context}}
\begin{tabular}{@{}p{1.75cm}p{2.35cm}p{2.35cm}@{}}
\toprule
Property & Competitive data & Coursework data \\
\midrule
Authors & Experienced, self-selected participants & Novice students in a shared curriculum \\
Tasks & Algorithmic problems with room for personal scaffolding & Constrained assignments, often tied to the same examples and interfaces \\
Volume & Multiple substantial solutions for active authors & Fewer and shorter submissions per author \\
Stability & More mature programming habits across rounds & Style may evolve across tasks and semesters \\
Shared code & Limited common code beyond problem format & Specifications, templates, skeletons, and instructor examples may overlap \\
Labels & Contest account usually maps to one participant & Submission labels may be affected by collaboration, copying, or outside help \\
\bottomrule
\end{tabular}
\end{table}

This distinction matters because source code attribution is not only a model-selection problem. A classifier needs repeated, author-associated regularities that remain visible after task-specific tokens and solution logic are controlled for. Competitive benchmarks often provide many such regularities: personal input-output wrappers, debugging idioms, preferred data structures, and repeated decomposition patterns. Coursework can suppress the same signal when all students are solving a small number of highly similar exercises, especially if the assignment design encourages a narrow set of acceptable solutions.

The comparison also clarifies the role of the open-assignment dataset. Open projects should, in principle, give students more freedom to express individual design habits. Their poor full-pool result is therefore not explained only by over-constrained task statements. It suggests that author-associated signal in educational settings can remain weak even when assignment form is less rigid, because novice style development, changing languages, small per-author history, and imperfect labels still operate together.

\section{Methodology}
We evaluate a primary CodeBERT pipeline and use a companion benchmark as secondary cross-model evidence. The goal is not to prove that modeling choices are irrelevant, but to test whether the competition-coursework gap remains visible under the evaluated architectures and task-holdout protocol.
\subsection{Models}
Our primary model is CodeBERT (\texttt{microsoft/\allowbreak codebert-base}) [1], a RoBERTa-style transformer pre-trained on paired natural language and source code. We attach a linear classification head with one output per candidate author and fine-tune the full network for multi-class sequence classification. Inputs are tokenized with the native CodeBERT tokenizer, truncated from the end when necessary, and padded to 512 tokens. Training uses the AdamW optimizer with a learning rate of $2\times 10^{-5}$, weight decay 0.01, batch size 8 per device, and 3-5 epochs depending on configuration, on dual NVIDIA T4 GPUs. All primary CodeBERT experiments use the same preprocessing and training implementation: the Kaggle GCJ scaling sets, the curated single-round GCJ sets, and the coursework datasets.

As architecturally different evidence, we report a character-level convolutional network (char-CNN) [12] trained from scratch for up to 12-14 epochs with model selection on validation accuracy. The companion benchmark reported in Section~5.4 uses the same task-holdout idea and evaluates a broader set of model families [17]. Alongside the character-level CNN it includes token-level convolutional networks, recurrent models in both unidirectional (LSTM) and bidirectional (BiLSTM) form, and transformer encoders used both as fine-tuned classifiers and as fixed feature extractors with mean pooling. The benchmark uses common split definitions and evaluation metrics, while model-specific preprocessing necessarily differs across architectures.

\subsection{Evaluation Protocol}
Evaluation uses a task-holdout protocol. For each author, the available tasks are partitioned so that the tasks used for testing are disjoint from those used for training. Every author is therefore evaluated only on solutions to tasks the model never saw that author solve, which reduces the risk that the classifier exploits task-specific patterns instead of author-associated style. The primary experiments use a fixed split for each configuration; authors with too few retained tasks are excluded. Comments are stripped where they might leak author names. We report Top-1 and Top-10 accuracy, following the practical scenario in which the system proposes a shortlist of candidate authors to an instructor rather than a single verdict. The Top-10 value for the 10-author setting is therefore reported only for completeness, since the correct author must be inside a ten-candidate shortlist. Random-guess Top-1 baselines ($1/N$ for $N$ authors) are provided for reference.

The companion benchmark additionally reports chance-adjusted Top-1 accuracy and proxy verification ROC-AUC [17]. Chance adjustment subtracts the random-guess baseline from observed Top-1, which helps compare candidate sets of different sizes. The proxy verification task asks whether two fragments share an author rather than naming the author outright; this formulation is closer to educational review practice, where the relevant question is often whether a submission is compatible with a declared student's previous work rather than which student should be accused. These companion metrics are interpreted as secondary evidence rather than as part of the primary CodeBERT comparison.

\subsection{Interpretive Framing}
The primary experiments are closed-set identification experiments: every test sample is assumed to have been written by one of the candidate authors known to the model. This is a standard formulation in authorship-attribution research, but it is stronger than the situation faced in teaching practice. In a real course, an instructor often asks a narrower question: whether a submitted program is consistent with a student's own previous work, whether it resembles another student's work, or whether it deserves closer human review. A closed-set classifier can still be informative for such workflows, but only if its Top-$k$ shortlist and confidence behavior are reliable enough to reduce rather than create investigative noise.

For that reason, we interpret Top-1 accuracy conservatively. High Top-1 performance on a clean benchmark shows that a model can recover author-associated regularities under that benchmark's assumptions. Low Top-1 performance on coursework does not prove that no authorship signal exists, but it does show that the evaluated pipeline cannot recover a useful signal at the full candidate-pool scale. Top-10 accuracy is included because a shortlist is closer to how an instructor might use a tool; however, a useful shortlist still needs to be much smaller and more accurate than random review, especially when the consequences of a false suspicion are serious.

The companion verification and abstention analyses are therefore important as qualitative checks. Verification asks a more deployable same-author question, while abstention tests whether the model can identify cases where it is sufficiently confident to be useful. In the companion benchmark, contest data provide such a high-confidence region, but the full coursework datasets do not. This makes the negative result more relevant than raw Top-1 alone: the weakness is not merely that the model ranks the correct author second or third, but that confidence does not recover a reliable operating regime on the most realistic coursework configurations.

Finally, all results are interpreted at the dataset-configuration level. The companion benchmark contains 26 configurations, but several are derived from related source corpora by changing author count, language, project size, or per-author file count. They are useful for triangulating mechanisms, not for claiming 26 independent replications. This distinction is important for the manuscript's evidential logic: the primary contribution is the controlled transfer test with CodeBERT, while the companion benchmark checks whether the same direction of evidence survives changes in model family and dataset configuration.

\section{Results}
We organize the results around the four research questions. We first establish how accuracy scales with the number of candidate authors on competitive data (RQ1), then test whether that accuracy transfers to institutional coursework (RQ2), check whether the observed behavior is specific to a single architecture (RQ3), and finally interpret dataset properties that may explain the performance gap (RQ4).
\subsection{Scaling on Competitive Programming Data (RQ1)}
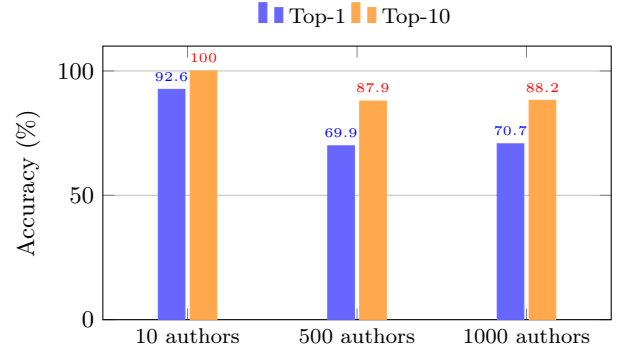
\begin{figure}[t]
\centering
\begin{tikzpicture}
\begin{axis}[
  ybar, bar width=10pt, width=\linewidth, height=5.2cm,
  ymin=0, ymax=110, ylabel={Accuracy (\%)},
  symbolic x coords={10 authors,500 authors,1000 authors},
  xtick=data, x tick label style={font=\small},
  legend style={at={(0.5,1.03)},anchor=south,legend columns=2,draw=none,font=\small},
  nodes near coords, nodes near coords style={font=\tiny},
  enlarge x limits=0.25, ymajorgrids, tick align=inside,
]
\addplot+[fill=blue!60,draw=blue!60] coordinates {(10 authors,92.6)(500 authors,69.9)(1000 authors,70.7)};
\addplot+[fill=orange!70,draw=orange!70] coordinates {(10 authors,100.0)(500 authors,87.9)(1000 authors,88.2)};
\legend{Top-1, Top-10}
\end{axis}
\end{tikzpicture}
\caption{CodeBERT accuracy on multi-round GCJ data as the number of candidate authors grows.\label{fig:scaling}}
\end{figure}

Table~\ref{tab:gcj} and Figure~\ref{fig:scaling} summarize CodeBERT performance on the multi-round GCJ configurations. With 10 authors the model reaches 92.6\% Top-1, and Top-10 accuracy is trivially 100\% since only ten candidates exist. Scaling to 500 and 1000 authors reduces Top-1 accuracy to about 70\%, while Top-10 accuracy remains near 88\%. Given random baselines of 0.2\% and 0.1\% respectively, the model performs far above chance and reliably narrows 1000 candidates down to a shortlist of ten.

\begin{table*}[t]
\centering
\caption{CodeBERT on GCJ configurations. Accuracy values are percentages.\label{tab:gcj}}
\begin{tabular}{lccccc}
\toprule
Configuration & Authors & Epochs & Top-1 (\%) & Top-10 (\%) & Random Top-1 (\%) \\
\midrule
GCJ multi-round, most active & 10   & 4 & 92.6 & 100.0 & 10.0 \\
GCJ multi-round, most active & 500  & 3 & 69.9 & 87.9  & 0.20 \\
GCJ multi-round, most active & 1000 & 5 & 70.7 & 88.2  & 0.10 \\
GCJ 2022, 10 tasks/author    & 636  & 5 & 27.9 & 51.2  & 0.16 \\
GCJ 2018, 10 tasks/author    & 1000 & 5 & 11.3 & 36.5  & 0.10 \\
\bottomrule
\end{tabular}
\end{table*}

\begin{figure}[t]
\centering
\begin{tikzpicture}
\begin{axis}[
  ybar, bar width=9pt, width=\linewidth, height=5.4cm,
  ymin=0, ymax=100, ylabel={Accuracy (\%)},
  symbolic x coords={{multi-round\\1000 auth.},{GCJ2022\\636 auth.},{GCJ2018\\1000 auth.}},
  xtick=data, x tick label style={font=\small,align=center},
  legend style={at={(0.5,1.03)},anchor=south,legend columns=2,draw=none,font=\small},
  nodes near coords, nodes near coords style={font=\tiny},
  enlarge x limits=0.3, ymajorgrids, tick align=inside,
]
\addplot+[fill=blue!60,draw=blue!60] coordinates {({multi-round\\1000 auth.},70.7)({GCJ2022\\636 auth.},27.9)({GCJ2018\\1000 auth.},11.3)};
\addplot+[fill=orange!70,draw=orange!70] coordinates {({multi-round\\1000 auth.},88.2)({GCJ2022\\636 auth.},51.2)({GCJ2018\\1000 auth.},36.5)};
\legend{Top-1, Top-10}
\end{axis}
\end{tikzpicture}
\caption{CodeBERT accuracy on multi-round versus single-round ($\approx$10 solutions per author) GCJ configurations.\label{fig:scarcity}}
\end{figure}
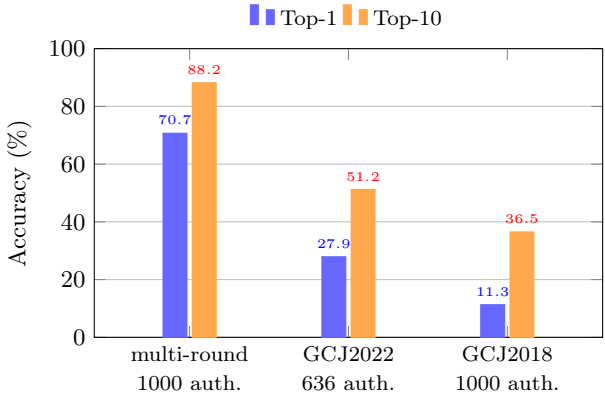

The lower half of Table~\ref{tab:gcj} and Figure~\ref{fig:scarcity} show CodeBERT accuracy on the single-round GCJ configurations, where each author has only about ten solutions. Here Top-1 accuracy is 27.9\% for GCJ 2022 (636 authors) and 11.3\% for GCJ 2018 (1000 authors), with Top-10 accuracy of 51.2\% and 36.5\% respectively. Compared with the multi-round setting, where the 1000 most active authors reach 70.7\% Top-1, this is a substantial drop, even though the task is still competitive programming. The comparison suggests that previously reported high GCJ accuracy depends partly on the availability of many solutions per author, and declines once each author is represented by only a small number of files. We note one caveat: the multi-round and single-round configurations also differ in provenance, since the former come from the public Kaggle repository and the latter from a curated GCJ archive, so differences in curation, author selection, and contest years may contribute to the gap alongside per-author volume.

\subsection{Limited Transfer to Institutional Coursework (RQ2)}
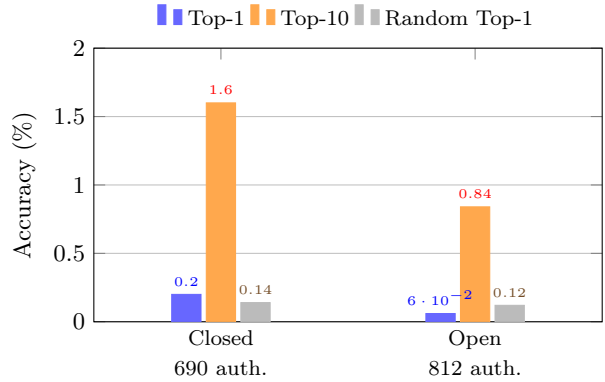
\begin{figure}[t]
\centering
\begin{tikzpicture}
\begin{axis}[
  ybar, bar width=11pt, width=\linewidth, height=5.2cm,
  ymin=0, ymax=2.0, ylabel={Accuracy (\%)},
  symbolic x coords={{Closed\\690 auth.},{Open\\812 auth.}},
  xtick=data, x tick label style={font=\small,align=center},
  legend style={at={(0.5,1.03)},anchor=south,legend columns=3,draw=none,font=\small},
  nodes near coords, nodes near coords style={font=\tiny},
  enlarge x limits=0.5, ymajorgrids, tick align=inside,
]
\addplot+[fill=blue!60,draw=blue!60] coordinates {({Closed\\690 auth.},0.2)({Open\\812 auth.},0.06)};
\addplot+[fill=orange!70,draw=orange!70] coordinates {({Closed\\690 auth.},1.6)({Open\\812 auth.},0.84)};
\addplot+[fill=gray!55,draw=gray!55] coordinates {({Closed\\690 auth.},0.14)({Open\\812 auth.},0.12)};
\legend{Top-1, Top-10, Random Top-1}
\end{axis}
\end{tikzpicture}
\caption{CodeBERT on the full-scale coursework datasets compared with random Top-1 baselines.\label{fig:student}}
\end{figure}

Table~\ref{tab:student} and Figure~\ref{fig:student} report the same CodeBERT pipeline applied to the coursework datasets. On the closed-assignment dataset of 690 authors, where every task is fully specified, Top-1 accuracy reaches only 0.2\%, close to the 0.14\% random baseline, and validation loss increases monotonically during training. This pattern is consistent with memorization of training files rather than transferable author-associated signal.

\begin{table}[t]
\centering
\caption{CodeBERT on student coursework. Values are percentages; both full-scale datasets are near or below their random baselines.\label{tab:student}}
\small
\setlength{\tabcolsep}{4pt}
\begin{tabular}{lccc}
\toprule
Dataset & Authors & Top-1 (\%) & Top-10 (\%) \\
\midrule
Closed & 690 & 0.2  & 1.6  \\
Open   & 812 & 0.06 & 0.84 \\
\bottomrule
\end{tabular}
\end{table}

The open-assignment dataset provides an important check because students design their own projects and have greater freedom of implementation. Evaluated over all 812 authors, however, the model reaches only 0.06\% Top-1 and 0.84\% Top-10, both below the corresponding random baselines (0.12\% and 1.23\%). An earlier run on a reduced pool of 100 authors had reached 11.3\% Top-1; the full-scale result suggests that this signal does not scale to the complete candidate pool. As on the closed dataset, validation loss rises monotonically throughout training, which is consistent with memorization rather than stable cross-task attribution.

\subsection{Cross-Model Evidence (RQ3)}
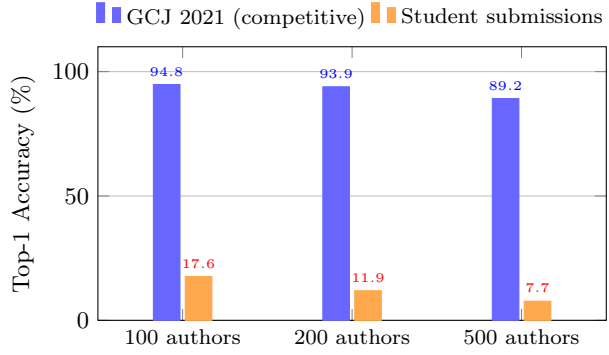
\begin{figure}[t]
\centering
\begin{tikzpicture}
\begin{axis}[
  ybar, bar width=10pt, width=\linewidth, height=5.2cm,
  ymin=0, ymax=110, ylabel={Top-1 Accuracy (\%)},
  symbolic x coords={100 authors,200 authors,500 authors},
  xtick=data, x tick label style={font=\small},
  legend style={at={(0.5,1.03)},anchor=south,legend columns=2,draw=none,font=\small},
  nodes near coords, nodes near coords style={font=\tiny},
  enlarge x limits=0.25, ymajorgrids, tick align=inside,
]
\addplot+[fill=blue!60,draw=blue!60] coordinates {(100 authors,94.8)(200 authors,93.9)(500 authors,89.2)};
\addplot+[fill=orange!70,draw=orange!70] coordinates {(100 authors,17.6)(200 authors,11.9)(500 authors,7.7)};
\legend{GCJ 2021 (competitive), Student submissions}
\end{axis}
\end{tikzpicture}
\caption{Char-CNN Top-1 accuracy on size-matched competitive (GCJ 2021) and student subsets.\label{fig:charcnn}}
\end{figure}

Figure~\ref{fig:charcnn} compares the char-CNN baseline on size-matched subsets of both families. On GCJ 2021 the char-CNN reaches 94.8\%, 93.9\%, and 89.2\% Top-1 accuracy for 100, 200, and 500 authors; on the coursework subsets of identical sizes it achieves only 17.6\%, 11.9\%, and 7.7\%. In the companion benchmark (Table~\ref{tab:benchmark}) the same qualitative pattern appears across additional configurations: competitive programming subsets retain high attribution accuracy, while coursework-derived configurations show substantially lower Top-1 accuracy, with the two full assignment sets near their chance baselines [17]. Increasing the number of files per student from 5 to 15 raises Top-1 accuracy only from 4.3\% to 7.8\%, whereas the analogous increase on GCJ 2020, from 5 to 13 solutions per author, lifts it from 53.5\% to 93.7\%. These results indicate that the observed performance gap is not specific to CodeBERT and persists across the evaluated model families.

\subsection{Companion Extended Benchmark Across Dataset Dimensions}
The following analysis is secondary evidence drawn from the companion benchmark [17]. Its role is to assess whether the pattern observed in the primary CodeBERT experiments persists across additional dataset configurations and model families, not to serve as an independent replication. The benchmark applies a task-holdout protocol to 26 evaluation configurations spanning contest, synthetic, and coursework-derived settings, and reports the best-performing model for each configuration. Table~\ref{tab:benchmark} lists representative configurations, while Table~\ref{tab:family} summarizes ranges by data context.

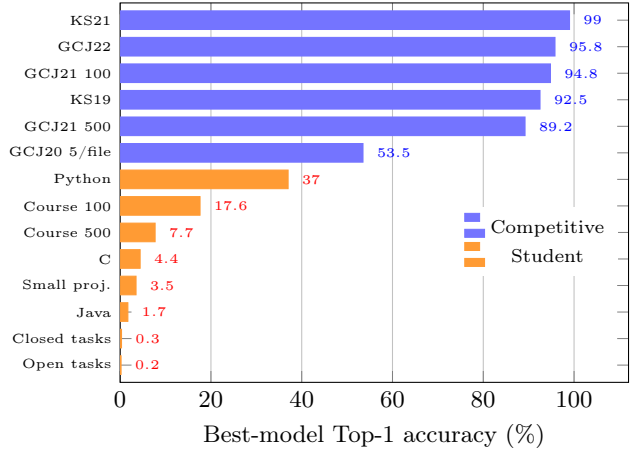
\begin{figure}[!htbp]
\centering
\begin{tikzpicture}
\begin{axis}[
  xbar, bar width=7pt, width=\linewidth, height=6.6cm,
  xmin=0, xmax=112, xlabel={Best-model Top-1 accuracy (\%)},
  symbolic y coords={Open,Closed,Java,Small proj.,C,CW 500,CW 100,Python,GCJ20-5,GCJ21-500,KS19,GCJ21-100,GCJ22,KS21},
  ytick={Open,Closed,Java,Small proj.,C,CW 500,CW 100,Python,GCJ20-5,GCJ21-500,KS19,GCJ21-100,GCJ22,KS21},
  yticklabels={Open tasks,Closed tasks,Java,Small proj.,C,Course 500,Course 100,Python,GCJ20 5/file,GCJ21 500,KS19,GCJ21 100,GCJ22,KS21},
  y tick label style={font=\tiny},
  nodes near coords, nodes near coords style={font=\tiny},
  every node near coord/.append style={xshift=2pt},
  enlarge y limits=0.05, xmajorgrids, tick align=inside,
  bar shift=0pt,
  legend style={at={(0.98,0.37)},anchor=east,draw=none,font=\scriptsize},
]

\addplot+[fill=blue!55,draw=blue!55] coordinates {
  (99.0,KS21)(95.8,GCJ22)(94.8,GCJ21-100)(92.5,KS19)(89.2,GCJ21-500)(53.5,GCJ20-5)};

\addplot+[fill=orange!80,draw=orange!80] coordinates {
  (37.0,Python)(17.6,CW 100)(7.7,CW 500)(4.4,C)(3.5,Small proj.)(1.7,Java)(0.3,Closed)(0.2,Open)};
\legend{Competitive, Student}
\end{axis}
\end{tikzpicture}
\caption{Representative companion-benchmark Top-1 accuracy [17].}
\label{fig:benchbar}
\end{figure}

The companion evidence is consistent with the primary CodeBERT comparison, but it also shows that the coursework side is not uniform. The two full assignment datasets approach the chance baseline, whereas some smaller or language-specific coursework subsets retain limited author-associated signal: the Python course reaches 37.0\% Top-1, and the 100-author coursework subset reaches 17.6\% Top-1 and 0.80 verification ROC-AUC. These values remain far below the corresponding contest configurations, but they should not be described as chance-level results.

\par\smallskip
\noindent\begin{minipage}{\columnwidth}
\centering
\scriptsize
\setlength{\tabcolsep}{3pt}
\captionof{table}{Selected companion configurations from [17]. Values are percentages except AUC.\label{tab:benchmark}}
\begin{tabular}{@{}p{1.45cm}p{2.35cm}rrrr@{}}
\toprule
Context & Configuration & Authors & T1 & T10 & AUC \\
\midrule
Contest & KS 2021 & 100 & 99.0 & 99.6 & 1.00 \\
Contest & GCJ 2022 & 100 & 95.8 & 98.8 & 1.00 \\
Contest & GCJ21, 500 auth. & 500 & 89.2 & 95.9 & 1.00 \\
Contest & GCJ20, 5 files/auth. & 210 & 53.5 & 72.2 & 0.93 \\
Synthetic & Control corpus & 100 & 43.0 & 85.7 & 0.96 \\
Coursework & Python course & 500 & 37.0 & 62.7 & 0.93 \\
Coursework & 100 authors & 100 & 17.6 & 46.9 & 0.80 \\
Coursework & 500 authors & 500 & 7.7 & 24.0 & 0.78 \\
Coursework & C course & 500 & 4.4 & 18.8 & 0.73 \\
Coursework & Java course & 97 & 1.7 & 10.0 & 0.49 \\
Assign. & Closed tasks & 690 & 0.3 & 1.5 & 0.50 \\
Assign. & Open projects & 812 & 0.2 & 0.9 & 0.50 \\
\bottomrule
\end{tabular}
\end{minipage}
\par\smallskip

\par\smallskip
\noindent\begin{minipage}{\columnwidth}
\centering
\footnotesize
\captionof{table}{Context-level ranges in the companion benchmark [17].\label{tab:family}}
\setlength{\tabcolsep}{4pt}
\begin{tabular}{@{}p{2.4cm}rrrr@{}}
\toprule
Context & Cfg. & T1 range & T10 range & AUC range \\
\midrule
Contest & 12 & 53.5--99.0 & 72.2--99.6 & 0.93--1.00 \\
Synthetic & 1 & 43.0 & 85.7 & 0.96 \\
Coursework subsets & 11 & 1.7--37.0 & 6.6--62.7 & 0.47--0.93 \\
Full assignments & 2 & 0.2--0.3 & 0.9--1.5 & 0.50 \\
\bottomrule
\end{tabular}
\end{minipage}
\par\smallskip

\par\smallskip
\noindent\begin{minipage}{\columnwidth}
\centering
\small
\setlength{\tabcolsep}{4pt}
\captionof{table}{Training-length sensitivity averaged over the companion configurations [17]. Model checkpoints are selected by validation performance.}
\label{tab:epochs}
\begin{tabular}{llrrrr}
\toprule
Model & Ep. & T1 & T5 & T10 & AUC \\
\midrule
char-CNN & 5 & 43.1 & 49.8 & 53.2 & 0.80 \\
char-CNN & 8 & 45.8 & 53.3 & 57.3 & 0.82 \\
char-CNN & 12 & 47.4 & 54.3 & 58.2 & 0.83 \\
transf. & 3 & 41.4 & 49.4 & 53.6 & 0.81 \\
transf. & 5 & 45.2 & 53.4 & 57.5 & 0.83 \\
\bottomrule
\end{tabular}
\end{minipage}
\par\smallskip

Several dimensions narrow the plausible explanations. The pre- and post-LLM-period GCJ configurations remain comparable, so the observed gap is not explained by the calendar period alone. Per-author volume matters within contest data: increasing GCJ 2020 from five to thirteen solutions per author raises Top-1 accuracy from 53.5\% to 93.7\%. The analogous coursework comparison, from five to fifteen files per author, raises Top-1 only from 4.3\% to 7.8\%. Language and project-size partitions show the same pattern: Python coursework is the strongest student subset but remains below the contest configurations, while C, Java, and project-size subsets remain much lower. Together, these results suggest that dataset and task characteristics impose a stronger limitation than the choice among the evaluated architectures.

A confidence-based abstention analysis in the same benchmark sharpens this interpretation. When the model is allowed to answer only on its most confident cases, contest configurations approach near-perfect retained accuracy, indicating a usable high-confidence operating region. Coursework configurations benefit much less, and the closed assignment set does not recover useful accuracy under abstention. This asymmetry matters for deployment, because an integrity tool must know when to remain silent, and the examined coursework data provide little evidence of a reliable confidence region.

\subsection{Dataset Properties and Plausible Explanations for the Performance Gap (RQ4)}
Based on the observed dataset characteristics and prior literature, we identify several factors that may contribute to the lower performance on coursework data. These factors should be interpreted as plausible mechanisms rather than independently established causal effects.

\noindent\textit{Directly documented dataset properties.}
\begin{itemize}
  \item \textbf{Author skill and style maturity.} GCJ datasets contain code from experienced programmers, whereas the coursework corpus consists mostly of early-study programming assignments. Developing command of a language can produce more uniform, textbook-like code with fewer stable idiosyncratic habits.
  \item \textbf{Task size and complexity.} Contest tasks are algorithmic problems that often contain recurring personal scaffolding and substantial implementations. Course assignments are shorter and more constrained, limiting the number of informative features available within a 512-token window.
  \item \textbf{Shared templates and specifications.} Students commonly work from the same task statements, examples, expected interfaces, and in some cases templates. This raises between-author similarity and weakens the contrast that a classifier needs.
  \item \textbf{Task- and language-dependent profiles.} A student's code may vary across C/C++ and Java assignments, and across open projects versus constrained exercises. This can split one experimental label into several apparent styles.
\end{itemize}

\noindent\textit{Plausible sources of additional variability or label noise.}
\begin{itemize}
  \item \textbf{Style evolution during study.} Students actively develop their programming practices during a course sequence, so code written at different stages may not express a stable cross-task profile.
  \item \textbf{Deliberate style modification.} Some students may alter naming, structure, or formatting when they know submissions are inspected, but this study does not measure such behavior directly.
  \item \textbf{Collaboration and copying.} Coursework labels are based on submission records, not independently verified authorship. Collaboration, copying, or external assistance can weaken the correspondence between the label and the actual source of the code.
  \item \textbf{Different revision conditions.} Contest solutions are produced under time limits, whereas coursework may be rewritten, formatted, or edited over a longer period. The direction and size of this effect are not isolated here, but it plausibly changes the observed signal.
\end{itemize}

These mechanisms help interpret the internal ordering of the results without turning the study into a causal analysis. The single-round GCJ results (Section~5.1) show that per-author code volume accounts for part of the gap: with only about ten solutions per author, Top-1 accuracy on competitive data already falls to 11-28\%. The remaining difference between those GCJ configurations and the full assignment datasets is consistent with the documented coursework constraints above, especially shared tasks, short submissions, developing student profiles, and imperfect labels.

\section{Discussion}
Taken together, the results show that the strong attribution accuracy reported on competitive programming data should not be read as a general capability of deep attribution models. The same CodeBERT pipeline that separates a thousand Google Code Jam programmers performs at or below chance baselines on the examined full-scale coursework datasets, and the companion benchmark shows that this gap persists across the evaluated model families. The most defensible interpretation is therefore narrower than a universal claim about source code authorship: contest data expose a stronger and more repeatable author-associated signal than the examined coursework data under the evaluated protocols.

The practical implication for academic integrity tooling is direct. Accuracy figures published on GCJ must not be used by themselves to justify deployment on coursework submissions. In the examined institutional setting, full-scale closed and open assignment datasets do not support reliable closed-set attribution. If used at all, attribution should serve only as a weak screening signal in a Top-$k$ shortlist or verification-support workflow with mandatory human review, and never as sole evidence in an academic misconduct case [17].

This does not make competitive-programming data useless. GCJ remains valuable for controlled experiments because many authors solve shared problems and because large author pools can be constructed without exposing student records. Its weakness is ecological rather than technical: it is a favorable benchmark for measuring whether a representation can exploit clear author-associated signal, but it is not a proxy for the institutional setting in which attribution evidence would be used. A stronger benchmarking practice would therefore report not only accuracy, but also author-selection criteria, tasks per author, files per author, average fragment length, held-out task definitions, and the degree of shared scaffolding. Without those details, two datasets with the same number of authors can represent substantially different attribution problems.

The results also argue against treating model architecture as the main bottleneck. CodeBERT and the char-CNN differ strongly in their inductive biases: one relies on a pre-trained contextual token representation, the other learns character patterns from scratch. The companion benchmark extends this comparison to token CNNs, recurrent networks, and transformer encoders. None of these evaluated choices removes the gap between contest and coursework configurations. A better model may still improve the absolute numbers, but the more important question is whether the input data contain enough stable, author-associated evidence for the intended decision. In the examined coursework setting, final source files alone appear to be a weak evidence source.

The companion benchmark broadens the comparison from two primary coursework datasets to a wider map of dataset configurations. It does not show that coursework code is universally unidentifiable, but it does show that the examined student-derived configurations contain substantially less transferable author-associated signal than contest configurations. Some reduced coursework subsets retain limited signal, especially in Python and smaller candidate pools, whereas the two full assignment datasets approach chance. This points to a dataset- and task-formulation problem under the evaluated modeling approaches, and suggests that progress in education is likely to require richer contextual evidence such as repository-visible process behavior, code-quality profiles, calibrated verification, and retrieval tools that help instructors inspect the relevant context rather than stronger classifiers applied only to the final artifact [17, 18].

An educational system built from these findings should therefore start from review support rather than accusation. The relevant user is not a classifier that must output a single author, but an instructor who needs to compare a submission with a student's previous work, inspect similar code fragments, and decide whether there is a pedagogically meaningful reason to ask follow-up questions. In that setting, attribution scores should be calibrated, accompanied by nearest-neighbor examples or repository-level evidence, and explicitly marked as non-conclusive. The privacy requirements are also different from public contest analysis: coursework data contain reused educational materials and student records, so the raw submissions cannot be treated as an open benchmark. In this study, student and repository identifiers were pseudonymized, no personally identifiable information is reported, and the institutional data are available only under appropriate restrictions from the corresponding author.

The negative result is also useful for assessment design. If instructors want authorship-style evidence to be meaningful, they need longitudinal and diverse programming histories: multiple tasks, multiple contexts, and enough opportunities for stable personal habits to emerge. Very small tasks, heavily scaffolded exercises, and sudden language switches are good for teaching specific concepts, but they are poor material for attribution. This does not mean such assignments should be changed to help classifiers. It means that any integrity workflow relying on style must account for the educational purpose and structure of the assignments from which the evidence is drawn.

\section{Limitations and Threats to Validity}
Several limitations qualify our findings. First, the coursework corpus originates from a single institution and a specific set of courses; other curricula may induce more or less stylistic diversity. Second, the multi-round GCJ scaling configurations deliberately select the most active authors, creating a favorable high-volume benchmark. The single-round GCJ configurations partly bridge this by fixing about ten solutions per author, but a perfectly matched comparison is impossible because the contest and coursework populations differ in more than one dimension. Third, hold-out evaluation within one course context may still overestimate real-world deployment performance, since deployment would involve code written months after the training data, possibly after a student's style has evolved. Fourth, hyperparameters were tuned modestly; more aggressive optimization or substantially different representations could improve coursework performance, so our claim is limited to the evaluated CodeBERT pipeline and companion model families.

Fifth, author labels in coursework are based on submission records and may be affected by collaboration, copying, or external assistance, placing an upper bound on measurable attribution accuracy. This is not only a measurement limitation, but also an ethical constraint on interpretation: a mismatch between a submission and a model's learned representation of a student cannot by itself establish authorship, originality, intent, or misconduct. Sixth, the raw coursework submissions cannot be released publicly because they contain educational materials reused across course editions; this limits external reproducibility even though the anonymized institutional data can be requested under institutional restrictions. A further threat is provenance: the multi-round competitive data and the curated single-round GCJ archive come from different repositories, so curation differences may contribute to part of the competitive-side variation. Finally, the mechanisms discussed in Section~5.5 are plausible explanations grounded in observed dataset properties and prior work, not independently randomized causal tests.

\section{Conclusions}
This study asked whether the strong published performance of deep learning-based source code authorship attribution transfers from competitive programming to institutional coursework. Using the same primary CodeBERT pipeline evaluated with a task-holdout protocol and Top-$k$ metrics, we confirmed high performance on multi-round Google Code Jam data (92.6\% Top-1 for 10 authors; 70.7\% Top-1 and 88.2\% Top-10 for 1000 authors), but observed at-or-below-baseline performance on the two full-scale coursework datasets: 0.2\% Top-1 on the closed-assignment dataset of 690 authors and 0.06\% Top-1 on the open-assignment dataset of 812 authors. A character-level CNN baseline showed the same direction on size-matched subsets (89-95\% on GCJ versus 8-18\% on coursework), and the companion benchmark provided consistent cross-model evidence across additional configurations (RQ1-RQ3).

The most plausible explanation is not a single failed model, but the combination of dataset and task characteristics in the examined coursework setting (RQ4). Documented properties include novice and developing author profiles, shared task specifications, shorter submissions, task-dependent style, and language-dependent variation. Additional label noise may arise from collaboration, copying, external assistance, or deliberate style modification, although those mechanisms are interpretive rather than directly isolated in this study. Open-ended projects are especially informative: despite greater design freedom, performance falls below the random Top-1 baseline when evaluated over the full pool of 812 authors, after appearing stronger only in a reduced 100-author setting.

We draw two recommendations. For researchers, GCJ should be treated as a favorable benchmark rather than evidence of direct transfer to coursework; claims of practical applicability should be validated on the target educational context, with Top-$k$ and verification metrics reported alongside Top-1. For educators, attribution models should currently be used, if at all, only as review-prioritization aids with mandatory human oversight, never as standalone evidence of misconduct. Future work should therefore move toward style-stability modeling across semesters, calibrated verification, and hybrid evidence that combines learned embeddings, repository-visible behavior, code-quality metadata, and instructor-facing retrieval over the student's programming history [17, 18].



\begin{thebibliography}{99}
\bibitem{codebert} Feng F., Guo X., Tang D., Duan N., Feng X., Gong M., et al., CodeBERT: A Pre-Trained Model for Programming and Natural Languages, In: Findings of the Association for Computational Linguistics: EMNLP 2020, 2020
\bibitem{caliskan} Caliskan-Islam A., Harang R., Liu A., Narayanan A., Voss C., Yamaguchi F., et al., De-anonymizing Programmers via Code Stylometry, In: Proceedings of the 24th USENIX Security Symposium, 2015
\bibitem{dlcais} Abuhamad M., AbuHmed T., Mohaisen A., Nyang D., Large-Scale and Language-Oblivious Code Authorship Identification, In: Proceedings of the 2018 ACM SIGSAC Conference on Computer and Communications Security (CCS), 2018
\bibitem{alsulami} Alsulami B., Dauber E., Harang R., Mancoridis S., Greenstadt R., Source Code Authorship Attribution Using Long Short-Term Memory Based Networks, In: Computer Security - ESORICS 2017, LNCS vol. 10492, 2017
\bibitem{kalgutkar} Kalgutkar V., Kaur R., Gonzalez H., Stakhanova N., Matyukhina A., Code Authorship Attribution: Methods and Challenges, ACM Computing Surveys, 2019, 52(1)
\bibitem{bogomolov} Bogomolov E., Kovalenko V., Rebryk Y., Bacchelli A., Bryksin T., Authorship Attribution of Source Code: A Language-Agnostic Approach and Applicability in Software Engineering, In: Proceedings of ESEC/FSE, 2021
\bibitem{dauber} Dauber E., Caliskan A., Harang R., Shearer G., Weisman M., Nelson F., et al., Git Blame Who? Stylistic Authorship Attribution of Small, Incomplete Source Code Fragments, Proceedings on Privacy Enhancing Technologies, 2019, 2019(3)
\bibitem{frantzeskou} Frantzeskou G., Stamatatos E., Gritzalis S., Katsikas S., Effective Identification of Source Code Authors Using Byte-Level Information, In: Proceedings of the 28th International Conference on Software Engineering (ICSE), 2006
\bibitem{burrows} Burrows S., Uitdenbogerd A.L., Turpin A., Comparing Techniques for Authorship Attribution of Source Code, Software: Practice and Experience, 2014, 44(1)
\bibitem{krsul} Krsul I., Spafford E.H., Authorship Analysis: Identifying the Author of a Program, Computers and Security, 1997, 16(3)
\bibitem{bert} Devlin J., Chang M.-W., Lee K., Toutanova K., BERT: Pre-training of Deep Bidirectional Transformers for Language Understanding, In: Proceedings of NAACL-HLT 2019, 2019
\bibitem{charcnn} Zhang X., Zhao J., LeCun Y., Character-level Convolutional Networks for Text Classification, In: Advances in Neural Information Processing Systems 28 (NeurIPS), 2015
\bibitem{vaswani} Vaswani A., Shazeer N., Parmar N., Uszkoreit J., Jones L., Gomez A.N., et al., Attention Is All You Need, In: Advances in Neural Information Processing Systems 30 (NeurIPS), 2017
\bibitem{archive} jur1cek, Google Code Jam dataset, Kaggle dataset, \url{https://www.kaggle.com/datasets/jur1cek/gcj-dataset} (accessed 2026); Google, Coding Competitions Archive: Code Jam and Kick Start, GitHub repository, \url{https://github.com/google/coding-competitions-archive} (accessed 2026)
\bibitem{horvathSurvey} Horv{\'a}th M., Pietrikov{\'a} E., Spinellis D., Bridging Behavioral Biometrics and Source Code Stylometry: A Survey of Programmer Attribution, arXiv preprint arXiv:2603.11150, 2026
\bibitem{horvathAIP} Horv{\'a}th M., Pietrikov{\'a} E., Gurb{\'a}\v{l} F., et al., Personalized Learning Analytics Through Static Code Analysis in Computer Science Education, Acta Informatica Pragensia, 2026, 15(1), 54--71, \url{https://doi.org/10.18267/j.aip.283}
\bibitem{horvathProcess} Horv{\'a}th M., Evaluating Static and Process Evidence for Code Authorship in Programming Education, arXiv preprint arXiv:2607.07400, 2026
\bibitem{horvathRetrieval} Horv{\'a}th M., Pietrikov{\'a} E., Evaluating Semantic and Quality-Aware Retrieval for Source Code Repositories, arXiv preprint arXiv:2607.09161, 2026
\end{thebibliography}
\end{document}